# Stochastic parallel gradient descent optimization based on decoupling of the software and hardware

Qiang Fu,[a,b,c,d] Jörg-Uwe Pott,[a] Feng Shen,[b,c] Changhui Rao[b,c], Xinyang Li[b,c]

[a] Max-Planck-Institut für Astronomie, Königstuhl 17, D-69117 Heidelberg, Germany

[b] The laboratory on Adaptive Optics, Institute of Optics and Electronics, Chinese Academy of Science, Chengdu, 610209, China

[c] The key laboratory on Adaptive Optics, Chinese Academy of Sciences, Chengdu, 610209, China

[d] Graduate School of Chinese Academy of Science, Beijing, 100039, China

fu@mpia.de

Abstract

We classified the decoupled stochastic parallel gradient descent (SPGD) optimization model into two different types: software and hardware decoupling methods. A kind of software decoupling method is then proposed and a kind of hardware decoupling method is also proposed depending on the Shack-Hartmann (S-H) sensor. Using the normal sensor to accelerate the convergence of algorithm, the hardware decoupling method seems a capable realization of decoupled method. Based on the numerical simulation for correction of phase distortion in atmospheric turbulence, our methods are analyzed and compared with basic SPGD model and also other decoupling models, on the aspects of different spatial resolutions, mismatched control channels and noise. The results show that the phase distortion can be compensated after tens iterations with a strong capacity of noise tolerance in our model.

## 1. Introduction

Many optical systems usually work in the stable environment to keep the high performance. When the stability is disrupted, they usually suffer the performance degradation due to the dynamic perturbation of external environment like the atmospheric turbulence. Thus, the perturbation needs to be removed to improve the performance with regard to the laser beam combination [7], optical imaging in telescope, et al. The active correction methods are usually used to correct the dynamic distortion. The dominating method is the wave-front conjugation correction thanks to the accurate measurement by the wave front sensor (WFS) and the key component deformable mirror (DM) in most cases as corrector. As higher spatial resolution of the imaging system is required, the actuators of DM need to be increased enormously. Estimation shows that the efficiency is lowered as $N^2$ when the control actuators number $N$ increased and the matrix computation involved in Wave-front Conjugation correction (WFC) is only efficient for low resolution (N<200-300)[4]. The direct substitution with high resolution device in the primary system is almost infeasible, while the advanced controlling method is necessary.

The other type of the active correction method is the model-free optimization, which is also named image sharpening correction method and is nearly discarded in the last century due to its low computation performance and heavy computation burden [9]. Nevertheless, with improvement of the computation capability of modern computers and the demand of the high resolution control, it is possible to reactivate this technology which has the advantage of simple structure without wave-front sensors. Several decades ago, the typical optimization algorithm was the climbing mountain algorithm [6] and currently turns to the stochastic parallel gradient descent (SPGD) optimization algorithm [7,8,12,13]. They have low convergence velocity since the normal performance metric referred to the light intensity is coupled into global control information such as metrics correlated to light intensity [14]. The convergence velocity of SPGD algorithm is reduced by $\sqrt{N}$ when the control channel $N$ increased [11].

A number of researchers have applied the SPGD algorithm successfully to many aspects like coherent beams combination [7], laser beam clean-up [27], atmospheric laser communications [28], et al, where the aberration usually changes slowly. However, very few people concentrate on the improvement of the algorithm performance to extend it to the more general condition. M.A. Vorontsov proposed a decoupled SPGD (DSPGD) [14, 16] algorithm incorporating wave-front senor aiming to decouple the performance metric to accelerate the convergence. However, the wave-front sensors based on interferometer is not easy to be realized and will make the system more complex. This may turn the merit of unnecessary WFS to the shortcoming. If and only if the radically enhanced performance can be gained, it is possible to introduce the WFS in SPGD model. In this paper, a simple decoupled method is reconsidered based on atmosphere turbulence without sensors, and also another decoupled method with novel S-H wave-front sensor as a slope sensor is proposed. These are the main concern of the improvement of SPGD algorithm in this paper. This may also be extended to other optimized evolving algorithms, such as genetic algorithm [30], simulated annealing algorithm [31], et al.

In section 2, we firstly classified the decoupled method into two different types, software and hardware decoupling. In software decoupling, the normal SPGD algorithm depending on the control of Zernike basis instead of voltages of corrector is considered as a decoupling way which is analyzed in a new point of view. In hardware decoupling, we then develop a new model which is delineated explicitly based on normal S-H sensor. In addition, all of the DSPGD control methods are analyzed based on low orders of Zernike aberration in this part. In section 3, the mismatched model between wave-front sensor and corrector related to the different control channels is analyzed in detail. In section 4, the noise tolerance is discussed. In section 5, on the base of numerical simulation, the DSPGD method is investigated through correcting atmospheric turbulence aberration on different spatial resolution(8×8, 16×16 and 32×32 control channels).

## 2. Development of decoupled SPGD optimization technique

## 2.1. Overview of both SPGD algorithm and original decoupled methods

Firstly, SPGD algorithm will be reviewed below. It is a model-free iteration control method, which is initialized in 1997 by M.A.Vorontsov [12]. The basic iteration equation is:

$$u^{n+1} = u^n - \gamma \delta J \delta u \quad (1)$$

*u* is the control vector of voltage which is applied on Deformable Mirror(DM). *r* is the spatial coordinate. *n* is the iteration number. $\gamma$ is the ration scale. *J* is the optimized target function and is also used to be the performance metric. $\delta J$ is the performance metric variation. $\delta u(r)$ is the perturbation voltage vector, which follows the Poisson random distribution or Gaussian random distribution on each iterative step, e.g. the probability density distribution $P(\delta u = \pm \tau) = 0.5$. $\gamma \delta J \delta u(r)$ is approximate to gradient ($-du/dt$) of control vector. There are many performance metrics which are commonly used for the specific applications.

$$J_1 = \frac{\iint \sqrt{(x-x')^2 + (y-y')^2} I(x,y)dxdy}{\iint I(x,y)dxdy} \quad (2)$$

$$J_2 = \iint I^2(x,y)dxdy \quad (3)$$

$$J_3 = \iint_R I(x,y)dxdy \quad (4)$$

$$J_4 = \frac{I_{\max f}(x,y)}{I_{th\max f}(x,y)} \quad (5)$$

*x'* and *y'* are the light intensity distribution centroid, *x* and *y* are the distribution coordinates of light intensity. *I(x,y)* is the light intensity on every pixel. $I_{maxf}$ is the experimental maximum light intensity of far field and $I_{thmaxf}$ is the theoretical maximum light intensity of far field. As far as we know, the mean square radius of metric $J_1$ is the most effective performance metric [29] since it combines the light intensity and location information. $J_2$, $J_3$ and $J_4$ are only referred to the entire light intensity or partial intensity. $J_4$ is also the definition of Strehl ration. $I_{\max} = \max\left(\iint F\{A\exp(-i\varphi)\}\right)^2$, where *F{}* is the symbol of Fourier transform operator; *max*() is the operator of gaining maximum value; *A* is the wave-front amplitude and $\varphi$ is the distortion phase distribution. For different applications, the choice of the performance metrics may be diverse, but all these performance metrics mentioned in this paper are all on the base of Strehl for convenience.

Although the convergence can be accelerated by selecting suitable performance metric, it still needs over hundreds of iterations [29]. The main cause of the slow velocity is the coupled performance metric. It is also analyzed by M.A.Vorontsov [14] who has put forward several general decoupled methods. Here, the concept is repeated and some different ideas are generated. Let's decouple the *J* in Eq.(1): $J = j_1, j_2..., j_n$; $j_n$ is corresponding to the DM actuator distribution. Then the iterative equation is

$$u^{n+1}(r) = u^n(r) - \gamma(\delta j_1, \delta j_2, ..., \delta j_n)\delta u(r) \ . \tag{6}$$

The metric variation $\delta j$ in Eq.(6) is defined in Eq.(4) and usually converges to minimum.

The advantage is that it can accelerate the convergence effectively whereas it makes the system more complex, since it needs new module such as interferometer. There is not a standard module like interferometer realized in the system up to now. So the goal that we want to achieve is to develop a most probable method based on the existing system to explore the decoupling algorithm.

**2.2. Software decoupled method**

If we only consider the decoupled metric in Eq.(6), the focus is thus to decompose the wave-front on an intelligent way. Because the wave-front can usually be decomposed by orthogonal Zernike basis or Karhunen-Loeve modes [1], the general idea is to look for the correlation between the orthogonal modes and the control vector.

When we consider the aberration correction of the atmospheric turbulence, there is an accelerated SPGD method called Model SPGD correction [13]. This method transforms the optimized voltage vector of corrector to the mode coefficients of wave-front Zernike basis without introducing any extra hardware. It could be defined as a soft decoupled correction method(SDC) while the method proposed in[14] could be defined as a decoupled correction method(HDC) with hardware. In SDC, *J* is the decoupled metric on the base of Zernike basis. The interested basis order depends on the number of DM actuator. $\gamma$ is the amplitude of $\delta u(r)$ which is usually a constant for each control channel. The ration of Zernike basis coefficient varies with different types of aberrations. For instance, in atmospheric turbulence which is affected by wind [2], the tip-tilt error and defocusing error take up over 80%. If the $j_n$, the ratio of each decoupled performance, can be adjusted according to the proportional turbulence Zernike coefficient, or in another speaking, the perturbation could vary with the Zernike coefficients，we can accelerate the convergence extremely. Fig.1 denotes the soft decoupled model. Theoretically, if we can decouple metric *J* for each channel, the convergence velocity is only limited by the slowest control channel. However, the complete decoupling is obviously too difficult to be achieved in reality. The partial decoupling usually can be achieved such as metric *$j_n$* above. This could be also considered as mismatched situation described in section 3.

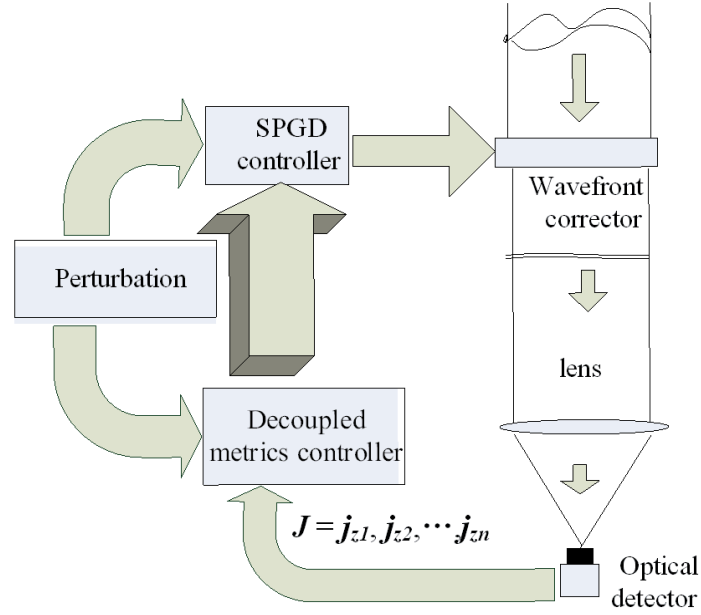

Figure 1. Soft decoupled SPGD model

SDC is generally difficult to be achieved since the ration of each Zernike basis is usually the statistical value which is difficult to be acquired in practice. In addition, we only optimize the low orders of Zernike coefficients that we are interested in and usually ignore the residual higher orders. The proportion $\rho_j^2$ of atmospheric turbulence parameters depending on the Zernike basis is listed in Table 1. The first term of Zernike modes is the piston error which can be approximated to the wave-front average phase and be usually omitted in the normal correction.

Table 1. Residual Error Coefficient $A_N$ and variances proportion $\rho_j^2$ based on Zernike basis for compensation of Kolmogorov Turbulence

| order | 2(tip) | 3(tilt) | 4(defocus) | 5 | 6 | 7 | 8 | 9 | 10 |
|---|---|---|---|---|---|---|---|---|---|
| $A_N$ | 0.582 | 0.134 | 0.111 | 0.088 | 0.065 | 0.059 | 0.053 | 0.046 | 0.040 |
| $\rho_j^2$ | 0.448 | 0.448 | 0.0232 | 0.0232 | 0.0232 | 0.0062 | 0.0062 | 0.0062 | 0.0062 |

Now the iterative equation correlated to Zernike mode coefficient is

$$u_z^{n+1}(r) = u_z^n(r) - \gamma \delta J_z \delta u_z(r), \tag{7}$$

where
$$\delta u_z = \delta u \bullet \rho_u \tag{8}$$

, $\rho_u$ is the weight mentioned by M.A. Vorontsov[13] and $\delta u_z$ is the new perturbation based on residual aberrations. The perturbation optimization is beyond the definition of SPGD algorithm mentioned above. We maintain the initial perturbation and then decouple the metric $J$ as $J_z = J \bullet \rho_u$. Now we can get

$$J_z = (j_{z1}, j_{z2}..., j_{zn}) \qquad (9)$$

Eq. (9) shows that the performance *J* is decoupled based on Zernike basis. Then we can write it as a new type:

$$\overline{\delta J_z} = \delta J \bullet (jr_1, jr_2, ..., jr_n) . \qquad (10)$$

$jr_n$ is the **$n_{th}$** order of Zernike basis proportion $\rho_j^2$ based on the atmospheric aberration (Table 1) and $\delta J$ is globally coupled metric which is a scalar. $\overline{\delta J_z}$ is a vector which is considered to be the ensemble of individual *jr*. The updated equation Eq.(7) turns out to be $u^{n+1}(r) = u^n(r) - \gamma \overline{\delta J_z} \delta u(r)$.

The distortions of phase correspond to the Kolmogorov turbulence model in the analysis. The power spectrum of phase fluctuations [19] is expressed by

$$G(q) = (0.023 / r_0^{5/3}) q^{-11/3} \qquad (11)$$

*q* is the spatial frequency and *$r_0$* is the Fried parameter( it is also the notable coherence length). All the corrected aberrations in this paper follow this spectrum. Then the software decoupled method will be analyzed formally in the following.

The variation $\delta J$ of the normal metric of the SPGD algorithm is expressed by $\delta J = J(\phi + \delta\phi) - J(\phi)$, where $J = \frac{1}{\psi} \int \phi^2 d^2 r$. Then

$$\delta J = \frac{1}{\psi} \int \langle \delta\phi^2 \rangle d^2 r + \frac{1}{\psi} \int \langle \delta\phi \bullet \phi \rangle d^2 r . \qquad (12)$$

$\psi$ is the aperture size of corrector and $\phi$ is the wave-front which is usually decomposed based on the Zernike mode coefficients $\rho_j^2$ in Table.1. The first term of Eq.(12) is fixed when current $\phi$ is known and the key point is to maximize the second term. If we consider the aberration based on the orthogonal Zernike basis, residual aberration $\phi$ and perturbation $\delta\phi$ can be written as below respectively.

$$\delta\phi = \sum_{i=1}^{N} \delta a_i Z_i , \quad \phi = \sum_{i=1}^{N} a_i Z_i \qquad (13)$$

$a_i$ is the Zernike coefficient and $Z_i$ is the Zernike basis. Here, the normal control vector consists of the voltage of each actuator, so any control vector should be transformed to be the voltage before it is sent to the corrector. In the Eq.(13), $a = Au$, where *u* is voltage vector and *A* is transform matrix. After perturbation is generated, in order to obtain control voltage, we should take inverse operator of *A*. Then $u = a * A^+$.

The perturbation of basic metric *J* in Ref [13] associated to uncompensated aberration is

$$\langle \delta J \rangle \approx 2\alpha_0 \sum_{j=1}^{M} \langle \rho_j^2 \rangle \qquad (14)$$

The first $M$ orders of aberration are assumed to be corrected. $\alpha_0$ is the correlation coefficient between wave-front perturbation and Zernike coefficient of turbulence, which is usually a constant and should be much less than 1 defined in Ref[13]. The purpose is to find the appropriate $\alpha_0$ to maximize $\langle \delta J \rangle$ while achieving the optimum correction at one iterative step. For simplification, the correlation between residual wave-front and perturbation is represented as

$$\eta = \sum_{i=1}^{M} \langle a_i \delta a_i \rangle. \tag{15a}$$

In Eq.(15a), $a_i$ is the mode coefficient of the current turbulence and $\delta a_i$ is the current perturbation. In the SDC correction, the convergent velocity is accelerated with maximizing correlation coefficient between perturbation and residual aberration based on Zernike modes. Then for normalization, Eq.(15a) can be rewritten as:

$$\eta_E = \frac{\sum_{j=1}^{M} \langle b_j a_j \rangle}{\sqrt{\sum_{j=1}^{M} \langle |b_j| \rangle^2} \sqrt{\sum_{j=1}^{M} \langle a_j \rangle^2}} \tag{15b}$$

$\eta_E$ is redefined as correlation coefficient in Eq.(15b). The convergent velocity achieves optimum when $\eta$ attains maximum. $a_j$ in Eq.(15b) is identical with that in Eq.15(a). $b_j$ is the statistics perturbation corresponding to statistics variance of atmospheric turbulence. $b_j$ is the substitution of $\delta a_i$ in Eq(15a) for the general definition.

The number of optimized coefficient $M$ is usually smaller than the number of control channels. We can also optimize the interested modes when the wave-front phase is well cognized. Even though the number of modes equates to that of channels, the improvement is also obvious due to the increasing of the correlation between wave-front and perturbation. The explicit analysis will be conducted in the 3ed section.

The theoretical analysis from Eq.(12) to Eq.(15) explains the benefit of this method by some degree. However, it may confuse the definition of SPGD. From Eq.(7) to Eq.(10), we should take this into account and then $b_j$ should be redefined as $jr$ in Eq.(10). Now, it is assumed that the amplitudes of the perturbed coefficients of Zernike basis are transformed to be the amplitudes of the decoupled performance metric. In another speaking, the performance metric is decoupled by the proportional Zernike basis. Then, the approximate gradient of each channel is obtained as $\gamma \delta J \delta u(r)$. In this case, instead of equation Eq.(15b) we have

$$\eta_{NE} = \frac{\sum_{j=1}^{M} \langle jr_j a_j \rangle}{\sqrt{\sum_{j=1}^{M} \langle |jr_j| \rangle^2} \sqrt{\sum_{j=1}^{M} \langle a_j \rangle^2}} \tag{15c}.$$

This is also the reason of the definition of software decoupled method which is different from the work of M.A.Vorontsov [12].

**2.3. Hardware decoupled method**

The HDC method depicted in Fig.3(a) is on the base of wave-front sensors. The general wave-front sensors are the interferometers including point diffraction interferometer (PDI) and Zernike phase contrast interferometer [14] et al. However, these wave-front sensors are too complex to be realized in DSPGD model especially for atmospheric turbulence. Here, another useful scheme is described. In the WFC structure, the S-H sensor is a normal wave-front sensor in adaptive optics systems.

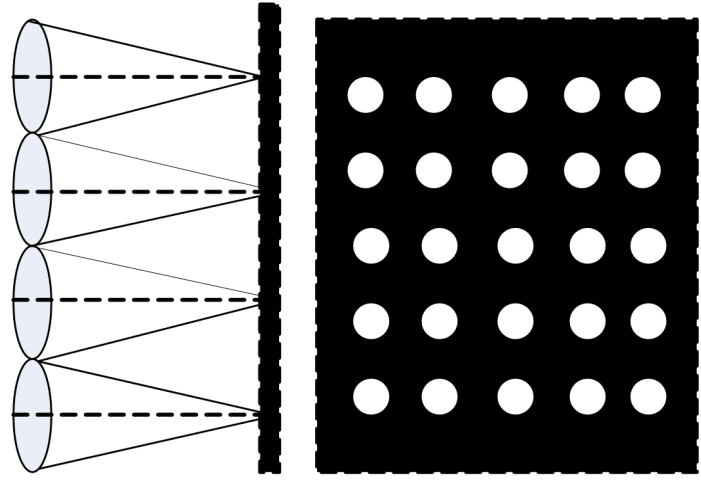

Figure 2. Shack-Hartman wave-front sensor

The principle of S-H sensor is repeated simply again in Fig. 2. The wave-front goes through micro lens array and then images on the focus. The gradient can be obtained through comparison between calibration and the real-time images. After that, the wave-front is rebuilt by the gradient information with different methods. For the typically modal correction, the equations are listed below:

$$S(x,y) = \sum_{i=1}^{N} A_i Z_i(x,y) \tag{16}$$

$$A_z = BU \tag{17}$$

$S(x,y)$ is the gradient distribution of incident wave-front. $i$ is the Zernike order. $N$ is the total number of Zernike orders. $A_i$ is the $i_{th}$ Zernike polynomial coefficient. $Z_i(x,y)$ is the gradient of $i_{th}$ Zernike polynomial basis. $U$ is the voltage vector of DM. $B$ is the influence function of actuator. When the gradient is obtained firstly, the Zernike coefficient matrix $A_z$ can be calculated from pseudo-inverse operation of $Z$. Thus, we get $A = S*[ZZ^+]^{-1}*Z^+$ from Eq.(16). Then the voltage can be calculated by $B$ on the same methods ($U = A_z * inv(B)$) where $inv()$ is the inverse operation. If we suppose that the total Zernike orders are 30 and the matrix of wave-front pixels size is 100×100, the size of matrix $A_i$ will be 1×30 and the size of matrix $Z$ will be 30×10000.

The computation budget will be very huge especially over the thousand actuators.

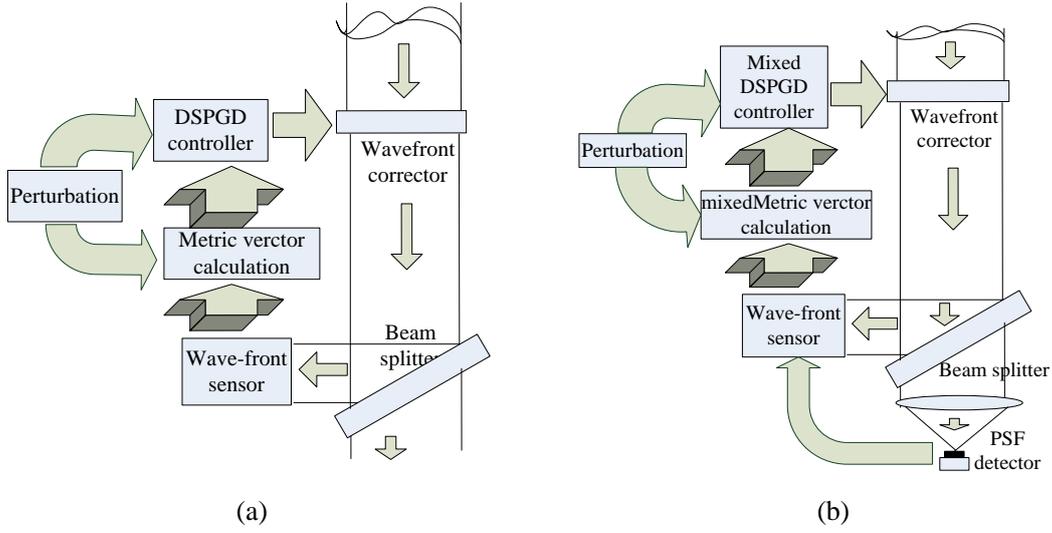

Figure 3. DSPGD models: (a) standard DSPGD model. (b) mixed DSPGD model.

The ultimate goal we want to achieve for correcting the aberration is to minimize the gradient of WFS, in another speaking, to flat the wave-front. Each sub-aperture gradient could be the performance metric applied to SPGD. This can decrease the computation burden effectively at the cost of reducing the close loop bandwidths. Actually the mass centroid of sub-aperture beam is only necessary on this method. Then the new metric is

$$S_r = \sqrt{(S_x - S_{x0})^2 + (S_y - S_{y0})^2} \tag{18}$$

The mass centroid is defined as $S_x = \dfrac{\sum xI(x)}{\sum I(x)}, S_y = \dfrac{\sum yI(y)}{\sum I(y)}$ in the Cartesian coordinate system. $S_x$ is the mass centroid of $x$ direction coordinate and $S_y$ is on the $y$ direction. $S_{x0}$ is the mass cetroid calibrated on the $x$ direction; $S_{y0}$ is calibrated on the $y$ direction; $S_r$ is the relative radius of beam position which will be the minimum after correction. The reason that we choose $S_r$ rather than metrics $J_1$, $J_2$, $J_3$ and $J_4$ described in the second section is that the image received by detector for each sub-aperture has fewer pixels and is insensitive to those metrics. Then，the Eq.(7) becomes

$$u^{n+1}(r) = u^n(r) - \gamma(\delta S_{r1}, \delta S_{r2}, ..., \delta S_{rn})\delta u(r). \tag{19}$$

The basic assumption is the identical control channels on both WFS and DM. However, in practice, the actuators number is usually less than the number of WFS sub-apertures. This is a kind of mismatched condition which will be specified in the third section. Even though the amount is equal, the scale is usually not matched. The mismatched scale between DM actuator and WFS subaperture has been analyzed based on continous surface DM in Ref [14].

**2.4. Mixed decoupling method based on hardware**

The number of continuous DM actuators is approximated to

$$N \approx (D/r_0)^2 \tag{20}$$

$D$ is the aperture of entrance pupil. The correction capability of DSPGD model also depends on both the actuator number of DM and sub-aperture number of sensor. The best condition is the identical control channels of both DM and corrected Zernike modes for turbulence. Then Eq.(20) becomes

$$N_{DM} = N_{modes} \approx (D/r_0)^2 \tag{21}$$

In the specific method proposed based on the S-H sensor, the normal metrics ($J_1$, $J_2$, $J_3$, $J_4$) cease to be effective since the light intensity in a single subaperture is not enough to offer the sensitive variation of the metric. For example, if there is a camera with 8 bit referred to the gray level of 0-255 scale, the one unit of the S-H sensor with 8×8 subapertures will only take the ratio 1/(8×8) of the entire light intensity of the pupil. The range of gray level would be also reduced by about tens times. It is easy to generate the idea to combine the far-field metric $J$ in Eq.(2) and S-H sensor metric in Eq.(18) since the global metric $J_2$ is also minimized like the slope in Eq.(18). The new iteration equation is stated below.

$$u^{n+1}(r) = u^n(r) - \gamma(\delta S_{r1}, \delta S_{r2}..., \delta S_{rm})\delta u(r) - \eta\gamma\delta J^n \delta u(r) \tag{22}$$

Eq.(22) shows the combination of far-field metric and near-field sensor metric in iterations. $\eta$ is the adjustable parameter which is usually smaller than 1 since the third term is the global metric and changes slowly. In addition, $\eta$ should be determined on certain conditions. We call this method the mixed DSPGD algorithm.

The mixed decoupling method makes use of the simple metric effectively which may accelerate the convergence potentially. However, there are problems probably when it is applied to practical system since both sensor output and the optical detector output may be out of sync where the metric doesn't fit to the theoretical expectation. It is supposed that the process is perfect synchronous in the discussion of this paper. In addition, another new parameter $\eta$ should be adjusted carefully.

### 2.5. Discussion of the decoupled methods

For the rough validity of the proposed methods above, the simple aberration is corrected by different methods in this part. The corrected aberration is the superposition of the first 10 orders of Zernike modes with the coefficients in Table.1. In the SDC, the first 5 orders is the optimized target. The corrector unit is typical distribution of symmetrical rectangle with 8×8 channels. The Strehl ration evolution of different methods for correcting the same initial aberration(Str≈0.27) is showed in Fig.4.

In order to simplify the definition, we define the soft decoupled method as DSPGD1, the HDC method with new performance metric in Eq.(18) as DSPGD2, and the HDC method with performance metric $J_2$ in Eq.(2) as DSPGD3. DSPGD4 is the mixed decoupling method with the iterative function in Eq.(22). Fig.4 shows the almost identical performance of both the soft decoupled method and the HDC method with new defined metric. The HDC method with metric $J_2$ only converges to the local extreme value, because the sub-aperture of WFS matched to detector only has few pixels which are not enough to build up the metric in Eq.(2). Fig. 4 also depicts that the DSPGD method converges to the extreme value after 20-30 iterations while SPGD method needs hundreds of iterations.

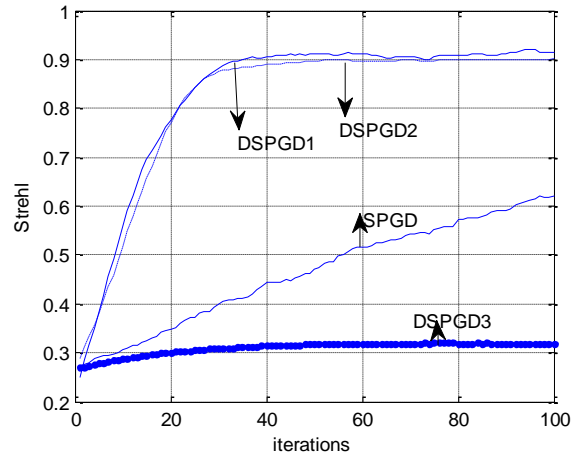

Figure 4. The different correction methods only for one order Zernike aberration with 8×8 units corrector, DSPGD1 is soft decoupled method, DSPGD2 is the HDC method with metric $S_r$ in Eq.(18), DSPGD3 is the HDC method with metric $J_2$ in Eq.(2).

The simple model is built up to compare the proposed method to the classical method in this part. Almost the identical performance could be achieved by our two proposed methods in this model and the great advantage of convergence velocity over the normal SPGD algorithm is also revealed. This model is useful especially for the aberration consisting of low order Zernike modes like laser purification [27].

## 3. Discussion of mismatched control channels in decoupled models

The complete decoupling means the elements numbers of both metric variation $\delta J$ and perturbation $\delta u(r)$ are equal. However, the WFS sub-apertures and DM actuators are usually not matched with each other perfectly. Firstly, one sub-aperture of sensor will be matched with more than one actuator. Secondly, more than one sub-aperture will be matched with one actuator of corrector.

$$\begin{cases} N(\delta J) < N(\delta u) \\ N(\delta J) > N(\delta u) \end{cases} \quad (23)$$

The mismatched situations are listed in Eq.(23). $N(\bullet)$ denotes the element number of the vector. The mismatched cases are obviously in HDC between sensors and correctors. There is still the mismatched situation in the SDC model without wave-front sensor. In SDC correction, it is inevitable to analyze the number of metrics which is more than the number of perturbed control channel since the orders of Zernike modes are usually less than actuator number of correctors[13].

**3.1. Mismatched control channels in SDC**

It has been analyzed that the DM with a certain structure has the certain correction capability when it is applied to atmospheric turbulence [2]. When the DM is fixed on the telescope pupil, the actuator spacing $d$ is related to $r_0$ which determines the 'fitting error' of DM. The analysis [2] shows that the corrected Zernike modes number should approximately equal to the actuators number. However, it is impossible to use the identical control channels with actuators in SDC since the performance has been improved through changing the big number of actuators channels to a small number of Zernike modes. The first 10 orders of aberrations are the key Zernike aberration and also the big scale aberration of turbulence [19]. It is feasible and meaningful to correct the concerned aberration by SDC method. The mixed perturbation applied in SPGD algorithm has been analyzed in Ref [13] to accelerate the convergence.

It is assumed that the first $N$ orders of Zernike aberration is compensated and the DM correction capability is the first $P$ orders of aberration where $P \geq N$. Then the residual turbulence aberration expectation becomes

$$\langle J_r \rangle = \sum_{j=N+1}^{P} \langle a_j^2 \rangle + \sum_{j=P+1}^{\infty} \langle a_j^2 \rangle \qquad (24)$$

The first term of Eq.(24) is the residual DM correction capability and the second term is the ultimately residual error after the DM reaching its limit. The residual errors of turbulence are $\sigma_N^2 = A_N \left( \frac{D}{r_0} \right)^{5/3}$ for $N \leq 10$ where $A_N$ is the fitting coefficient in Table 1 and $\sigma_N^2 \approx 0.2944 N^{-\sqrt{3}/2} \left( \frac{D}{r_0} \right)^{5/3}$ for $N \geq 10$[2] where $N$ is the corrected orders and $\sigma_N^2$ is the residual variance. During the whole evolution process, we can find that perturbation always exists. The perturbation is invariable if the scaling factor is fixed. Thus, compared to the WFS correction, this method would degrade the ultimate performance. When the convergence goes to the stability, the gradient will achieve the minimum, approximating zero, and hence the control vector will not vary any more. The components of gradient $\gamma$ and $\delta u_z(r)$ are the constants while the only variable is $\delta J_z$.

The adjustable parameters are the constants which should be adjusted carefully to keep the perfect performance. In practice, it is relatively difficult to adjust the parameter to be the optimum. Then an extra term should be added to Eq.(24):

$$\langle J_r \rangle = \sum_{j=1}^{N} \langle \chi_j^2 \rangle + \sum_{j=N+1}^{P} \langle a_j^2 \rangle + \sum_{j=P+1}^{\infty} \langle a_j^2 \rangle \qquad (25)$$

It is assumed that the first *N* orders are the optimized targets and the first *P* orders are the correction capability of the system in Eq.(25). The first term is the statistical variance of the first *N* orders of residual Zernike coefficients referred to perturbation $\delta J_z$ which can be considered as noise in the fourth section. Then the residual error is

$$\sigma_N^2 = \sigma_r^2 + A_N \left(\frac{D}{r_0}\right)^{5/3} = A_{new} \left(\frac{D}{r_0}\right)^{5/3},$$

where $\sigma_r^2$ is the residual variance of uncompensated Zernike aberration which are still in the correction range of DM, *D* is the aperture diameter of telescope and $A_{new}$ is the new fitting coefficient.

### 3.2. Mismatched control channels in HDC

The mismatched control channels are classified into 2 cases in HDC model stated in Eq.(23). The first case can be analyzed and realized by simple way. The actuators of DM are grouped to match the WFS. Each actuators group is treated as one unit and one individual perturbation voltage will be exerted on it. The grouped channels of sensor are depicted in Fig.5. The white dot on the black ground is the image of the sub-aperture of the sensor. The red dashed line divides the sub-apertures of sensor into 8×8 groups. Each group fits to the corresponding channel of corrector.

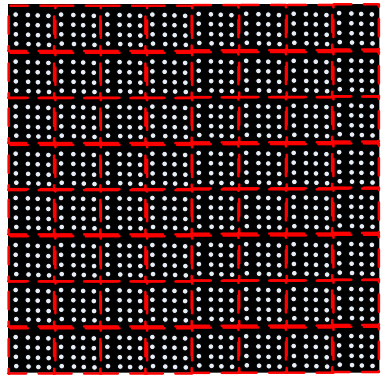

Figure 5. The structure of the 32×32 sub-apertures of S-H sensor which are divided by 8×8 sub-apertures

The first case could be considered to be the partial decoupling as the first equation of Eq.(23). In this case, we classified the actuators of DM into different groups matched to sub-apertures of wave-front sensor as described in Fig.3. The specific decomposing is

$$[\delta u_{11}, \delta u_{12},...\delta u_{1n}][\delta u_{21}, \delta u_{22},...\delta u_{2n}],...,[\delta u_{n1}, \delta u_{n2},...\delta u_{nn}] \quad (26)$$

*j* in Eq.(26) is a local metric, corresponding to actuators on the identical numbers, which will degrade the performance due to the local decoupling. There are 2 different decoupling methods in this case.

1. The sub-groups are combined to be an individual group applied on single perturbation for single group where any sub-group $[\delta u_{n1}, \delta u_{n2},...,\delta u_{nn}]$ can be shortened to be the $i_{th}$ element $\delta u_i$ of perturbed control vector in Eq.(26). Each sub-group is applied by the same perturbation. This method could decrease the resolution of the correctors whereas it is the complete decoupling.

2. The sub-group is considered to be the sub-system of SPGD where $[\delta u_1, \delta u_2,...,\delta u_n]$ could be a vector for different perturbations. This method is a partial decoupling since each sub-group could be considered as a individual SGPD system where a single metric obtained from the sensor will be matched with a single group. The individually iterative equation is $\overline{u}_{gi}^{n+1}(r) = \overline{u}_{gi}^{n}(r) - \gamma(\delta j_i)\delta\overline{u}_{gi}^{n}(r)$ where $\overline{u}_{gi}^{n}(r)$ is the updated $i_{th}$ group of control vector at $n_{th}$ step and $\delta j_i$ is the $i_{th}$ element of the metric obtained from the sensor. The actuators elements of each group are applied to the different perturbations. This method may increase the resolution of the first method at the cost of decreasing the convergence velocity.

The best method is the combination of these 2 conditions above if there is a unique corrector in the system. Then we get

$$\begin{aligned}u^{n+1}(r) = u^n(r) &- \gamma_1(\delta j_1, \delta j_2..., \delta j_n)(\delta u_1(r), \delta u_2(r),..., \delta u_n(r)) - \\ &\gamma_2(\delta j_1, \delta j_2..., \delta j_n)([\delta u_{11}(r), \delta u_{12}(r),..., \delta u_{1n}(r)]_1,...,[\delta u_{n1}(r), \delta u_{n2}(r),..., \delta u_{nn}(r)]_n)\end{aligned} \quad (27)$$

$\gamma_1$ and $\gamma_2$ are the adjustable parameters for the two methods in Eq.(27) respectively. Eq.(27) is something like Eq.(22) where the metrics of wave-front sensor and far-field detectors are combined. Eq.(27) will degenerate to Eq.(22) if the $(\delta j_1, \delta j_2..., \delta j_n)$ are combined together as $\delta J$ in the second term.

The new idea can be generated that normal SPGD algorithm could be accelerated based on the analysis above. Firstly, the wave-front with big-scale aberrations are corrected by the grouped actuators DM as low resolution corrector and then the small scaled aberration could be corrected by the high resolution corrector. This complicated model is called cascade adaptive optics system [8].

The second case is common, which we may encounter stated by the second equation of Eq.(23). Decomposing the whole metric *J* in Eq.(1) of normal SPGD algorithm, we then get

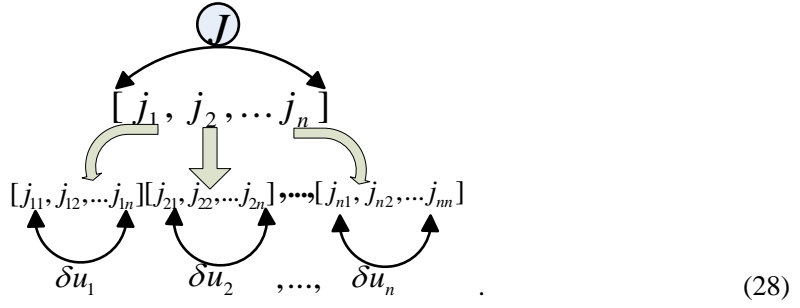

$$[j_1, j_2, \ldots j_n]$$
$$[j_{11}, j_{12}, \ldots j_{1n}][j_{21}, j_{22}, \ldots j_{2n}], \ldots, [j_{n1}, j_{n2}, \ldots j_{nn}]$$
$$\delta u_1 \quad \delta u_2 \quad, \ldots, \quad \delta u_n \quad . \tag{28}$$

Eq.(28) is the basic decomposition of metric applied in DSPGD model. The initial information obtained from the sensor is $j_{11}, j_{12}, \ldots j_{1n}, j_{21}, j_{22}, \ldots j_{2n}, \ldots j_{n1}, j_{n2}, \ldots j_{nn}$. Then they are divided by different group $j_n$ consisting of $j_{n2}, \ldots j_{nn}$. Finally, the separated metric $\boldsymbol{j_n}$ matches to the control channel $u_n$ of corrector. Comparing Eq.(28) to Eq.(26), we could find the difference between the number of decoupled metrics and the number of the control channels of corrector. Eq.(28) is the more common case since the sensor with the identical number of channels is low cost in manufacture than the corrector. The more accurate metric should be a weighted average since the grouped channels of practical wave-front sensor as described in Fig.5 is not matched with DM actuators exactly. The weighted average $j_n$ could be $j_n = a_{n1}j_{n1} + a_{n2}j_{n2} +, \ldots, a_{nn}j_{nn}$. $a_{nn}$ is the weighted coefficient depending on the actual architecture of wave-front sensor. The more complicated metric $\boldsymbol{J}$ should be considered only for non-interfering wave-front sensor.

However, the second case is a bit complicated while the performance metric obtained from the WFS should be averaged to fit to the channels of corrector. The difference appears in this case that the method with PDI is available to get the metric while the method with S-H sensor should adopt the new metric fitting to the new condition. The performance metric variation $\delta S_{ri}$ correlated to the WFS sub-apertures:

$$S_{ri} = \sqrt{S_{ip1}^2 + S_{ip2}^2 +, \ldots, + S_{ipn}^2} \tag{29}$$

$S_{ipn}$ is the slope of the $n_{th}$ sub-aperture of S-H sensor. $S_{ri}$ is the sum of all slopes of the sensor. The purpose is to minimize $S_{ri}$ to achieve the best Strehl ratio.

**3.3 simulation analysis for mismatched model**

Here, we still use the model as depicted in section 2.5 to analyze the mismatched condition. The channels of Corrector are expanded to 8×8, 16×16, and 32×32. The corrected aberration consisting of first 10 orders of the modes is identical from Fig.6 to Fig.8.

Fig.6 depicts the average Strehl convergent results with different resolution models. The convergence limits are almost identical since the corrected aberration is the low order with big scale. The Fig.7 depicts the performance comparisons of different mismatched conditions where the M-DSPGD1 stands for the first case in section 3.2 and M-DSPGD2 stands for the second case.

The convergence results of partial decoupling of mismatched model with Eq.(27) for two mismatched conditions in Fig.7(M-DSPGD1 and M-DSPGD2) show that the decreasing the resolution of DM can accelerate the convergence at the cost of lowering the convergence limit. The mismatched model in the second case shows that the partial decoupling in Fig.7(M-DSPGD2) lowers the convergence velocity apparently where it needs about 100 iterations to go to limit for low order aberration. This is still much better than normal SPGD model which needs over about 200 iterations to go to limit. The convergence limit of Strehl is only about 0.8 in Fig.7 because of the limitation of DM resolution.

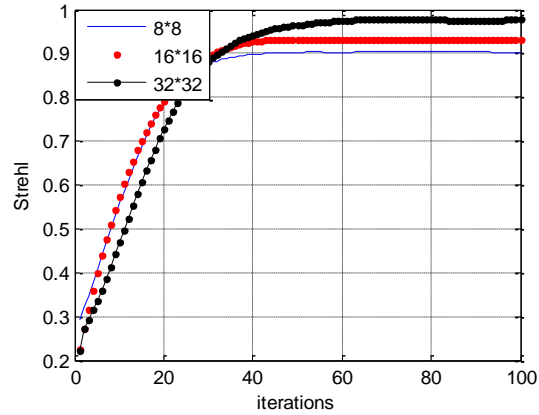

Figure 6. Evolution of Metric Strehl with different sub-apertures and matched control channels: 8×8, 16×16 and 32×32, the corrected aberration is a low order Zernike aberration( within first 5 orders).

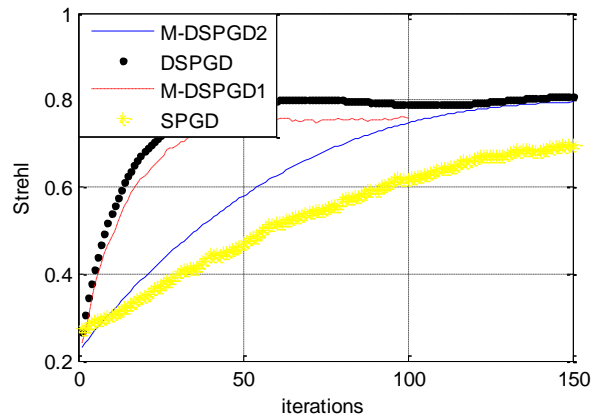

Figure 7. Evolution of Strehl with differently mismatched conditions. Black dot line is for DSPGD model with matched 8×8 units, red dash line(M-DSPGD1) is for mismatched model with 8×8 units sensor and 16×16 units DM where resolution of DM is lowered to 8×8 by grouped actuators, blue line (M-DSPGD2) is for mismatched model with 8×8 units sensor and 16×16 units DM, yellow line is for normal SPGD model of 8×8 units DM without wave-front sensor.

For the interfering wave-front sensor like PDI, $j_n$ could be expressed as $j_n = j_{n1} + j_{n2} +,...,+ j_{nn}$ which is the same to the second case in Section 3.2. The convergent velocity of DSPGD method with this metric can achieve the optimum compared to the matched structure since $j_n$ as the light intensity in Eq.(3) is sensitive to the phase changing. Fig.8 depicts performance of the hardware DSPGD method based on wave-front sensors differed in sub-aperture number and the corrector with the identical number of actuators. Different resolution models show the almost identical performance especially for iterations in Fig.8. There is slight difference among the convergence limits of three conditions due to the low resolution of simulation. We just use the simplified model to validate the analysis above, so the precision could be limited. The more explicit simulation is depicted in section 4.

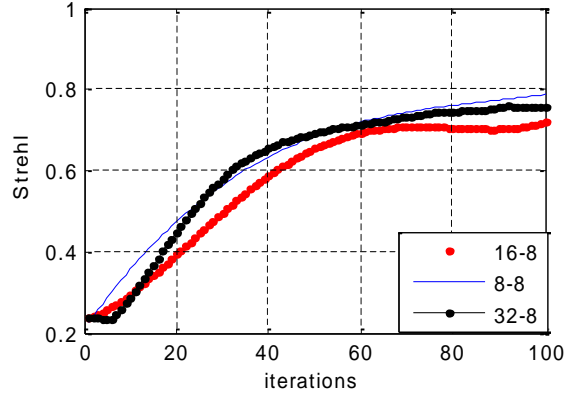

Figure 8. The averaged results of convergence based on different kinds of mismatched corrector and wave-front sensor. The corrector consists of 8×8 units and the wave-front sensors consist of 8×8 units (blue line), 16×16 units (red dot) and 32×32 units (black dot-line) respectively

## 4. The effect of system noise on DSPGD algorithm

In the WFS, it is inevitable to confront the vast majority of noise as analyzed in WFC [20, 24], but there are still some other problems which should be noted. Because SPGD algorithm is an iterative method, the random perturbation in Eq.(6) is approximated a part of the noise. Thus, the perturbation would be analyzed specifically in this part.

According to the source of perturbation, there are two types of SPGD algorithm [21]. One is the perturbation of algorithm on single direction and the other is the perturbation on double direction consisting of positive and negative parts in single iteration. Here two types of variation of metric $J$ in original Eq.(1) are:

$$\delta J^n = J^n - J^{n-1} \tag{30a}$$

$$\delta J^n = J^{n+} - J^{n-} \tag{30b}$$

Eq.(30a) shows the single direction which means the metric $J$ is obtained from the different iterative steps. The iterations in double directions in Eq.(30b) show that the metric $J$ is obtained from the individual iteration step. $J^{n+}$ and $J^{n-}$ denote the positive and negative perturbation on each iteration respectively. The iteration of algorithm on double directions is usually superior to that in the single direction since the former is not sensitive to the variation of performance metric including noise at the last step. So the perturbation in double directions can accelerate the convergence better than that in single direction [21].

In Eq.(6), there are 2 elements which are sensitive to the noise. One is the voltage $U(r)$ and the other is the performance metric $(j_1, j_2, ..., j_n)$. In the SPGD algorithm, the perturbed amplitude must be bigger than noise, otherwise the parameters will be contaminated resulting that the algorithm can't converge to the limit. Conversely, the perturbation could not be too larger to exceed the real gradient, otherwise leading to oscillation of correction all the time with large scale. Because the performance metric is decoupled to be small elements, the DSPGD model is sensitive to the noise than normal SPGD. The iterative equation could be written as follow.

$$u_{i+1}^{n+1}(r) = (u_i^n(r) + \Delta u_i) - \gamma(\delta j_1 + \Delta_{i1}, \delta j_2 + \Delta_{i2}, ..., \delta j_n + \Delta_{in})(\delta u(r) + \Delta \delta u(r)) \tag{31}$$

$\Delta u_i$ stands for the voltage noise; $\Delta j_{i1}$ is the performance noise and $\Delta \delta u(r)$ stands for the perturbation noise. We can extract the noise terms easily from Eq.(31). Thus, we get

$$\Omega = \Delta u_i - \gamma(\Delta j_{i1}, \Delta j_{i2}, ..., \Delta j_{in})\overline{\delta u}(r) - \gamma(\delta j_{i1}, \delta j_{i2}, ..., \delta j_{in})\Delta \delta u(r) . \tag{32}$$

$\Omega$ stands for the noise ensemble of all terms and could be also the noise of updated voltage $u_{i+1}^{n+1}(r)$. $\overline{\delta u}(r)$ is the perturbation contaminated by the noise. Then $\overline{\delta u}(r) = \delta u(r) + \Delta \delta u(r)$. Here, the analyzed noise is additive. The multiplicative noises are usually not taken into account for most detectors. The additive noise usually consists of readout noise, photos noise and dark current noise, et al. The perturbation noise $\Delta \delta u(r)$ could usually be neglected, since the noise is always the Gaussian type distributed like the perturbation. Their summation also fits to the Gauss distributed process. Then the noise of the third term in Eq.(32) could be a part of iterative process in Eq.(6). Eq.(33) can be shorten to be

$$\Omega = \Delta u_i - \gamma(\Delta j_{i1}, \Delta j_{i2}, ..., \Delta j_{in})\overline{\delta u}(r) . \tag{33}$$

Eq.(33) seems to be a iterative equation of DSPGD in Eq.(6). The difference is that the optimized target is the voltage noise $\Delta u_i$ in Eq.(33) and the noise is random on each iterative step. Eventually, the main noise sources which we should consider are the noise $\Delta j$ of the performance metric $J$ and voltage noise $\Delta u(r)$.

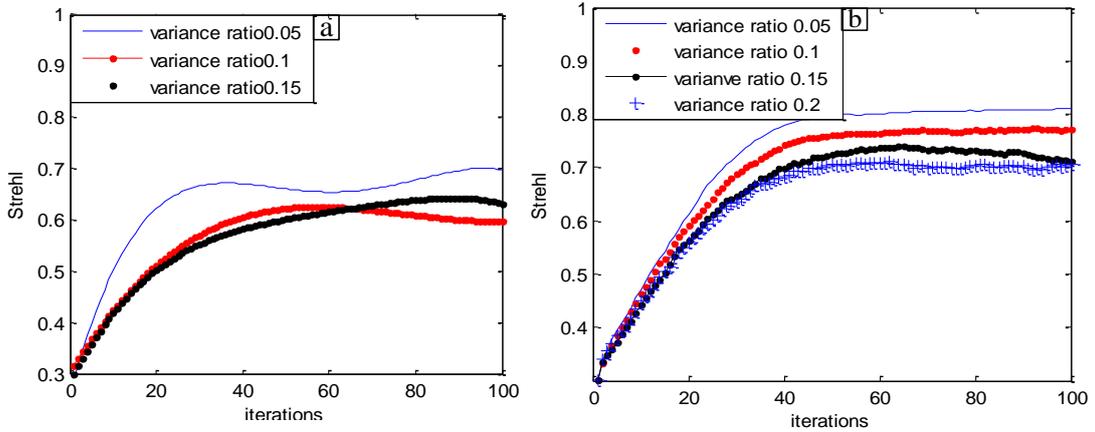

Figure 9. Strehl evolution for atmospheric turbulence affected by noise of different scale with the model described in the 2ed section. (a) noise impact on PDI sensor (b) noise impact on S-H sensor.

The calculated model is the same to that in section 2 and the corrector is $8 \times 8$ units matched to point diffraction interferometer. Fig.9 shows the sensitivity to the noise with different signal noise ratio (SNR). The noise ratio is defined by $R_{noise} = Var_{noise} / Var_{initial}$ where the *Var* is the symbol of variance. In the simulation, we found that the two main sources of the noise almost exerted the same effect on the model. Then the only one kind of noise left to be analyzed. The mean value of noise is 0 and RMS varies. In addition, the adjusting parameter $\gamma$ is reduced gradually followed by the noise growing larger. The phenomenon shows that the noise $\Delta j$ is a kind of perturbation if it fits to the initial perturbation type $\delta u(r)$. The convergence velocity becomes slow and the limit is also decreased with noise growing as showed in Fig.9. When $R_{noise}$ is bigger than 0.15, the ultimate Strehl is reduced by almost 10%. The decoupled method with S-H sensor is similar to that with interferometer sensor where the limit of convergence also degrades along with the increasing of the noise ratio. The comparison between Fig.9(a) and (b) shows that the decoupled method with S-H sensor has the bigger noise tolerance than the method with interferometer. For another speaking, the method with slope-type sensor is less sensitive to the noise than that with interferometer sensor. This can be generally accounted for by the different principle of the sensor used with which we obtain different metrics. Considering the general point-to-point interferometer, noise on each sampled point will directly affect the light intensity of each pixel. However, for the S-H sensor, only all the noise points in a subaperture affect indirectly a single slope.

## 5. Simulation analysis under atmospheric turbulence

### 5.1. Numerical model

We only analyze the type of continuous-deformation DM since almost all of the DM with high units (over 100) for correcting the atmospheric turbulence are continuous [3, 5]. The wave-front grid size is 128×128. The geometrical shape of the sub-apertures for both DM and WFS are identical. The soft decoupled SPGD model is described in section 3.A. The aberration of atmospheric turbulence is generated as the incident wave-front by the first 20ed Zernike basis with the statistical coefficients [2] of atmospheric turbulence. The HDC method incorporates both the typical point diffraction interferometer [16] and S-H sensor as the wave-front sensors. Each model is simulated by average of 50 phase screens. The configuration is the same to Fig.1. The Peak-Valley value of the incident wave-front phase of atmospheric turbulence is about 4.5 rad; the Root-Mean-Square (RMS) is about 1.2rad and the Strehl ration is around 0.27. The wave-front doesn't consist of the first (piston) and second (tip-tilt) Zernike aberration. All of the simulation work is on the base of the Matlab software of MathWorks Company. In addition, the mismatched corrector and sensor [14] are not analyzed.

### 5.2 Correction for static aberration

The soft decoupled correction (DSPGD1) method is shown in Fig.10(a) compared to the normal SPGD algorithm(red line). The convergence line of soft decoupled SPGD model with unoptimized perturbation ($\delta u$ in Eq.(7) with random distribution) is also depicted in Fig.10(a). Fig.10(b) shows that the improvement of correlation defined in Eq.15(c) between perturbation and residual wave-front can accelerate the convergence and that is also the reason of definition of soft-decoupled method(SDSPGD). Further more, the correlation coefficient of the normal SPGD model (red dot-line) is random showed in Fig.10(b). The correlated coefficient of normal SPGD is calculated by Eq.15(b) after transforming the perturbation $\delta u$ to the Zernike modes coefficients. The soft decoupled method is only implemented for the first 10 orders of Zernike aberration which are the main aberrations for atmosphere turbulence aberration showed in Table.1. The SDC method can also converge to the extreme value after 30 to 40 iterations. This requires that the statistic Zernike aberration components of wave-front should be learned previously.

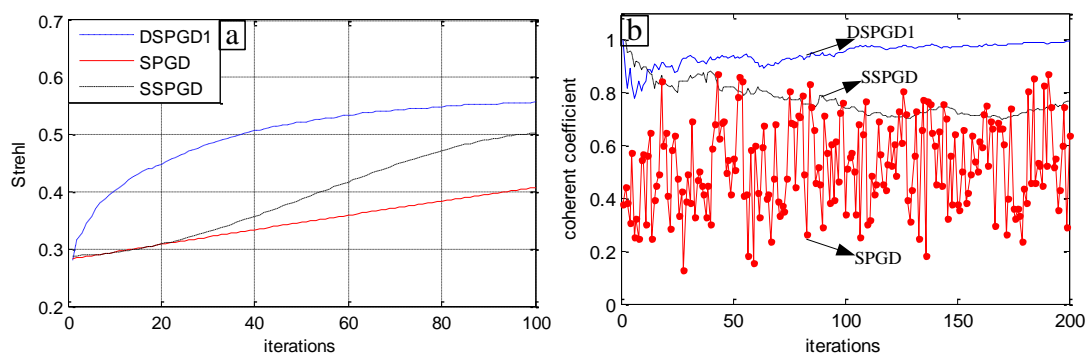

Figure 10. Comparison between SDC and normal SPGD optimized algorithm. (a) Strehl ration convergence of soft SPGD(SSPGD) without sensor, normal SPGD(red line) and soft decoupled SPGD(SDSPGD)(b) the evolution of correlation coefficients between perturbation $\delta u(r)$ and residual wave-front corresponding to (a).

The results in Fig.11(a) show that the $8\times8$ channels of low resolution corrector is insufficient to correct the aberration completely while the high resolution corrector with $16\times16$ and $32\times32$ units can compensate the distortion with the limit of Strehl ratio to over 0.9 on the same velocity only after 30-40 iterations. The comparison of different HDC methods is showed in Fig.11(b). For the complex aberration, the mixed decoupling method (red dot line) in the Fig.11(b) with S-H wave-front sensor is superior to the single S-H decoupled method(blue dashed line). The reason is that S-H sensor is not the point-to-point mapping sensor which could be sensitive to the noise with small scale. We have showed that the single low order aberration could be corrected based on S-H sensor as well as PDI sensor in Fig.4. For the turbulence aberration including more than the 10 orders of Zernike aberration, the gradient information is insensitive to the phase variation compared to the interferometer type sensor. The DSPGD model with S-H sensor is apt to trap in the local extrimum. In the simulation, we find all the repeated iterations with S-H sensor are trapped in the local extreme like Fig.11(b) denoted. The metric obtained from S-H sensor combined with far-filed Strehl could converge to global extremum in Fig11(b).

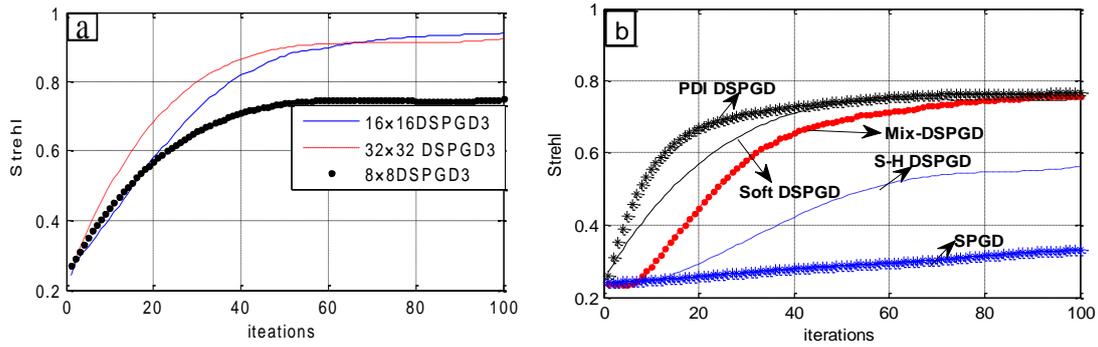

Figure 11. Comparison of different DSPGD methods with the identical wave-front. (a) the Strehl convergence of different spatial resolution corrector matched with point diffraction interferometer sensor on the identical wave-front based on decoupled method with hardware(b)comparison of different decoupled DSPGD methods including soft DSPGD(DSPGD1), hardware DSPGD (DSPGD2) with PDI, S-H sensor and mixed hardware DSPGD(DSPGD4) on the identical 8×8 channels.

The ultimate residual aberration after correction with mixed hardware methods (DSPGD4) is compared to that with PDI method(DSPGD2) in Fig.12. The S-H wave-front sensor is impossible to detect the piston type aberration [1], but the $2\pi$ ambiguity doesn't appear. The HDC method with S-H wave-front sensor is also impossible to correct piston aberration [14], but the $2\pi$ ambiguity exists in most cases as depicted in Fig.12(b). Two adjacent domains are super Owing to the selected metric slopes of the DSPGD4 method, the DSPGD4 method is able to suppress the $2\pi$ ambiguity since slopes is very sensitive to the variation of the controlling voltage.

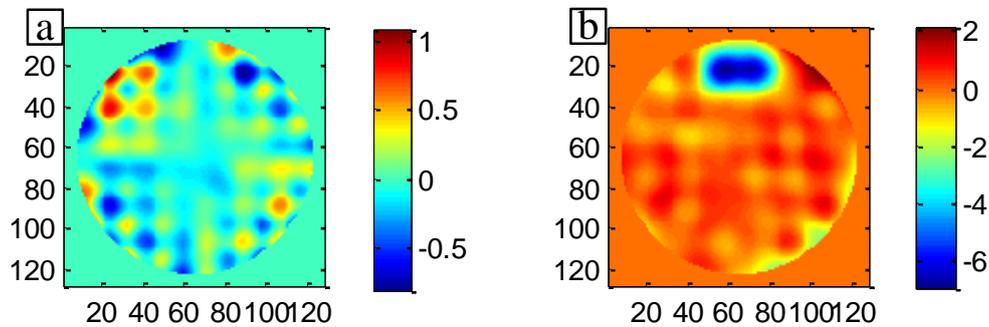

Figure 12. Comparisons of residual wave-fronts after correction of the identically incident wave-front. All coordinates of images are pixels and the Unit of the colorbar is rad. (a) DSPGD4 correction results with 8×8 units corrector (b) DSPGD2 correction results with 8×8 units corrector

The advantage of S-H wave-front sensor is that we can use S-H wave-front sensor or other slope-type sensor effectively under existing WFC system without designing any other complex device. Although the HDC method is a bit worse than the other decoupled methods, this method can correct the atmospheric aberration potentially with the easiest method to be implemented. The iterative correction method relies on the processing power of computation where the normal SPGD method has been realized by very large scale integration circuit (VLSI)[22,23]. It is supposed that a static aberration can be corrected at 30-50 iterations showed in this paper and the WFC correction method bandwidth is about 1k Hz. On the one hand, the situation that there are 3 times of sensor readouts time during each time of iteration of DSPGD model may decrease the bandwidth severely. On the other hand, there is no big matrix computation in DSPGD method. This can save a bit time. The ultimate bandwidth can be estimated about 1k/40 Hz on the identical computation ability for DSPGD correction model compared to the WFC correction.

**5.3 Correction for aberration affected by dynamic atmosphere turbulence**

The model-free optimization could be considered as an open loop control due to its relatively slow convergence rate which is implicitly performed on low wind velocity. On the contrary, if the capability of the device calculation could be improved, the shortcoming of the model would be overcome on some degree. The dynamic model of the atmospheric turbulence is build up according to the method in Ref.[25]. The first-order autoregressive model is introduced to describe the state equation: $\varphi_{n+1} = A\varphi_n + \sigma_n$ where $\varphi_n$ is the current state of atmospheric turbulence, $\varphi_{n+1}$ is the following state and $\sigma_n$ is the white noise of covariance matrix $C_v$. $C_v$ can be obtained by $C_v = AC_\varphi A^T - C_\varphi$ where $C_\varphi$ is the Zernike-basis covariance and $A^T$ is the transposition of the diagonal $A$ defined in Ref.[26]. Thanks to the modal cut-off frequency of the Power Spectral Density of a Taylor turbulence phase, the diagonal elements could approach to $a_i = \exp(-0.3(n+1)V/(f \cdot D))$ where $a_i$ is the $i_{th}$ diagonal element, $n$ is the radial order of Zernike basis, $V$ is the wind velocity, $f$ is the clock frequency and $D$ is the diameter of the telescope. Currently, the dynamic part is the wave-front $\varphi_n$ which is shifted by wind $V$. $D/r_0$ is used to characterize the atmosphere turbulence strength for the receiver system with the pupil $D$. Here, $D/r_0$ is set by classical ratio 6; $f$ is 1000Hz; $n$ is 11 (the amount order is reduced to 72)and $D$ is 0.8m.

The HDC method based on S-H sensor is compared to that based on PDI sensor for correcting the classical atmosphere turbulence with different wind velocity depicted above. The number of control channel is 8×8 and The subapertures of both sensors are 32×32. Therefore, the mismatched model in Eq.(28) is analyzed in this part. We investigated the achieved Strehl ratio after 100 iterations for different wind velocity on the same $D/r_0$.

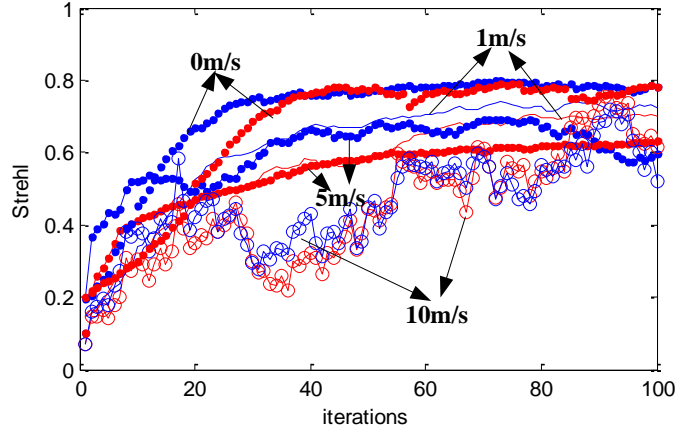

Figure 13. Correction for dynamic atmosphere turbulence on different wind velocities(1m/s, 5m/s and 10m/s). The HDC method based on PDI sensor is on the lines of blue type. The HDC based on S-H sensor is on the lines of red type.

The starting points of all curves in Fig.13 are random since each convergent process is conducted on the respective turbulence evolution $C_\varphi$ with the random process noise. When the wind velocity is increased, both convergent limits are decreased and the convergent processes become unstable in Fig.13. The worst processes fluctuating severely on the wind velocity 10*m/s* means that the wind velocity becomes unacceptable at this level. The convergence with S-H sensor is a little worse than that with PDI sensor since the method with PDI sensor only make use of the single perturbation given in Eq.(19), while the method with S-H sensor needs the mixed perturbation with two kinds of performance metrics given in Eq.(22). For the medium wind velocity less than 5 *m/s*, the 40-60 iterations will be needed to achieve the stable correction. Although the method with S-H sensor is not as good as that with PDI sensor, there is of interest to explore it in practice since the sensor is commonly used in traditional adaptive optics system. We have to indicate that the normal SPGD algorithm is not showed here since only the low wind velocity not more than 1*m/s* will radically impact the performance (1m/s leading to the lost convergence) at the same clock frequency to the above.

### 5.4. Discussion

We have compared the static and dynamic correction above between different methods. In the dynamic condition, the system seems not fast enough to correct the aberration in the medium turbulence. It is still necessary to state the algorithm in the other applications. Apart from the advantage stated, the shortcoming should be focused on as well. The main drawback is that the bandwidth is deduced severely when the extra sensor is introduced. This may be overcome when Very Large Scale Integrated Circuit(VLSI) substitutes the normal PC. Because the wave-front senor is added to the model-free method, the system model may be much more complex. In addition, the constraint of the specific sensors should be considered such as sensitivity to the noise.

Up to now, there is no standard sensor applied in decoupled method, so a set of sensors should be tested such as S-H sensor of the slope type sensor and interferometer type sensor. Moreover, the system correction capability should be estimated based on the WFC method. The comparison between WFC and model-free method is on the aspects of the control algorithm, system architecture, residual error, demand for Luminous flux, etc.

## 6. Conclusion

We mainly concern the improvement of the SPGD algorithm in this paper. The SDC method and HDC method are discussed explicitly. Based on hardware H-S sensor of the decoupled method, the wave-front conjugated correction is linked initially and efficiently to SPGD algorithm of model-free methods. The results of numerical simulation present that it's not as good as the completely decoupled SPGD method incorporating interferometer sensors, whereas it is superior to normal SPGD algorithm. The basic DSPGD method with S-H sensor is not available to achieve the best convergent limit. It should combine the far field metric or any other global metric obtained from the sensor to form a mixed DSPGD method achieving the high convergence velocity when the atmospheric turbulence aberration is corrected. In the current AO system, only using the algorithm without updating any hardware seems impossible to play a role in the stronger turbulence. However, there is still the potential prospect applied in the existing optical system when the slow variation of wave-front aberration appears. The noise impact on DSPGD method is also analyzed in the paper. The results show that the decoupled method with S-H sensor is more robust to the influence of noise than that with PDI sensor.

The soft decoupled method could be useful when the wave-front aberration mainly consists of the low order or big scale error which should be tested in real-time environment. For other applications, the hardware decoupled correction could be a potential consideration. In addition, the mismatched control channels could be met frequently for the existing systems which should be analyzed according to the methods in the paper. The experiment setup will be built up in the lab incorporating the S-H sensor to test the DSPGD method next.

## Acknowledgements

The study program is supported by the MPG-CAS graduate studentship.